\documentclass[a4paper, 11pt]{article}
\usepackage{epsfig}
\usepackage{amssymb}
\usepackage{amsthm}
\usepackage{lineno}
\usepackage{subfigure}
\usepackage[latin1]{inputenc}
\usepackage{graphicx}
\usepackage[top=1.5cm, bottom=2.0cm, left=2cm, right=1.5cm]{geometry}
\usepackage[english]{babel}
\usepackage{multicol}
\usepackage[absolute]{textpos}
\begin{document}
\title{Regional flow simulation in fractured aquifers using stress-dependent parameters}
\author{
    G. Preisig\thanks{Centre for Hydrogeology and Geothermics, University of Neuchâtel, Emile-Argand 11, 2000 Neuchâtel, Switzerland. giona.preisig@unine.ch}
   \and
   F.J. Cornaton
   \and
   P. Perrochet
}
\date{}
\maketitle
\begin{abstract}
A model function relating effective stress to fracture permeability is developed from Hooke's law, implemented in the tensorial form of Darcy's law, and used to evaluate discharge rates and pressure distributions at regional scales. The model takes into account elastic and statistical fracture parameters, and is able to simulate real stress-dependent permeabilities from laboratory to field studies. This modeling approach gains in phenomenology in comparison to the classical ones because the permeability tensors may vary in both strength and principal directions according to effective stresses. Moreover this method allows evaluation of the fracture porosity changes, which are then translated into consolidation of the medium.
\end{abstract}
\begin{center}
\textit{\small{numerical modeling, fracture hydraulic conductivity, effective stress, fractured rocks}}
\end{center}
\begin{multicols}{2} 
\section*{\large{Introduction}}
Crystalline and non karstic sedimentary rocks are anisotropic geological media with low hydraulic conductivity \cite{Neuman2005}. In such media, groundwater flow occurs primarily and sometimes exclusively through non-filled fractures. Their spatial arrangement (i.e. fracture network) leads to groundwater flow at a regional scale. At this scale, the most simple and useful way to conceptualize these aquifers is the equivalent porous media; the principal permeabilities of each fracture family are combined in space and result in a tensor describing the equivalent hydraulic conductivity of the rock mass \cite{Kiraly1969a, Berkowitz2002}.

The sensitivity of aquifers dynamics to effective stress was first described for granular porous media \cite{Terzaghi1923}; the process was then also observed in fractured aquifers \cite{Louis1969}. Nowadays, the dependency of fracture permeability on effective stress is a well known research topic and has been intensively studied during the last decades, especially to evaluate the stability of rock masses in presence of dams, tunnels, geologic radioactive waste repositories or CO2 sequestration fields \cite{Londe1987, Lombardi1988, Rutqvist2002, Zangerl2003, Ferronato2010}. In regional and deep groundwater flow systems, the reduction of water pressures leads to increasing effective stresses and decreasing permeabilities, with a possible consolidation of the aquifer. On the contrary, increasing groundwater pressures result in decreasing effective stresses and in increased permeabilities. Based on field and laboratory test results, Louis \cite{Louis1969} and later Walsh \cite{Walsh1981} derived respectively an exponential and a logarithmic model to explain permeability decreases with increasing effective stresses. The relationship between effective stress and permeability has been clearly identified both at local and regional scales via laboratory tests (see \cite{Tsang1981, Durham1997, Hopkins2000}), field tests (see \cite{Cappa2006, Schweisinger2009}), and observations of aquifer consolidation by measurements and modeling of ground subsidence (see \cite{Lombardi1988, Rutqvist1996, Zangerl2003}).

However, the equivalent porous medium approach and consequently the classical Darcy solution implemented in regional groundwater numerical models generally ignores this relationship \cite{Murdoch2006}, hence the interest in modifying the flow equation to a more realistic one, explicitly accounting for stress-dependent permeabilities. The present approach consists in inserting constitutive laws relating effective stress to permeability in the tensor form of Darcy's law, so that the permeabilities vary with stress (depth and geology) and water pressure. The constitutive model must: (1) respect most of the physical process at the microscopic scale but should also lend itself to practical application at large scales, and (2) be simple from a numerical point of view. Note that a number of rock mechanics codes exist (see for example \cite{Itasca2006, Abaqus2008, COMSOLMultiphysics2010, ZSOIL}) that solve coupled hydromechanical problems. However, these generally apply to relatively small scale problems, because they involve full and detailed deformation processes, and, therefore, become computationally prohibitive at hydrogeological scales. On the contrary, the present work focuses on a macroscopic approach allowing efficient large scale computations, while preserving the essence of the hydromechanical processes.

Firstly, a constitutive model is presented where fracture permeability is a function of the effective stress, as well as of the statistical distribution of the length of the asperities and their elasticity. Expressed in its tensor form, this law describes the process at the rock mass scale. Secondly, simulated stress-dependent permeabilities are compared with laboratory and field measurements from Durham \cite{Durham1997} and Cappa \cite{Cappa2006}. Thirdly, a finite element simulation is performed in order to illustrate this modeling approach. The constitutive model is also used to evaluate the changes in porosity between an initial and a modified hydrogeological state, and to compute the resulting subsidence.
\section*{\large{Constitutive aperture-stress model}}
The model considers a single fracture as a pair of surfaces, characterized by a set of asperities, the length of which follows a statistical distribution. This asperity population can be characterized by fracture morphology analysis (see \cite{Brown1995, Glover1998}). Assuming that each asperity $i$ obeys Hooke's law, the resulting normal stress, $\sigma_i$, proportional to its deformation is:
	\begin{equation}
		\sigma_i = E_i \frac{\Delta z_i}{z_i} = E_i \frac{z_i - a}{z_i} = \frac{F_i}{s_i}
		\label{eq:HookeLaw}
	\end{equation}
where the symbols stand for asperity original length $z_i$, compression $\Delta z_i=z_i-a$, elastic modulus $E_i$, average asperity section $s_i$, exerting force $F_i$ and fracture aperture $a$. Eq. (\ref{eq:HookeLaw}) implies the following conditions:
\begin{description}
\item[\textnormal{First,}] if $a \geq z_i$, $\sigma_i = 0$ (the asperity is at its original length).
\item[\textnormal{Second,}] if $a = 0$, $\sigma_i = E_i$ (the asperity is subjected to a total compression).
\end{description}
\begin{figure*}[!ht]
  \begin{center}
  	\includegraphics[width=0.75\textwidth]{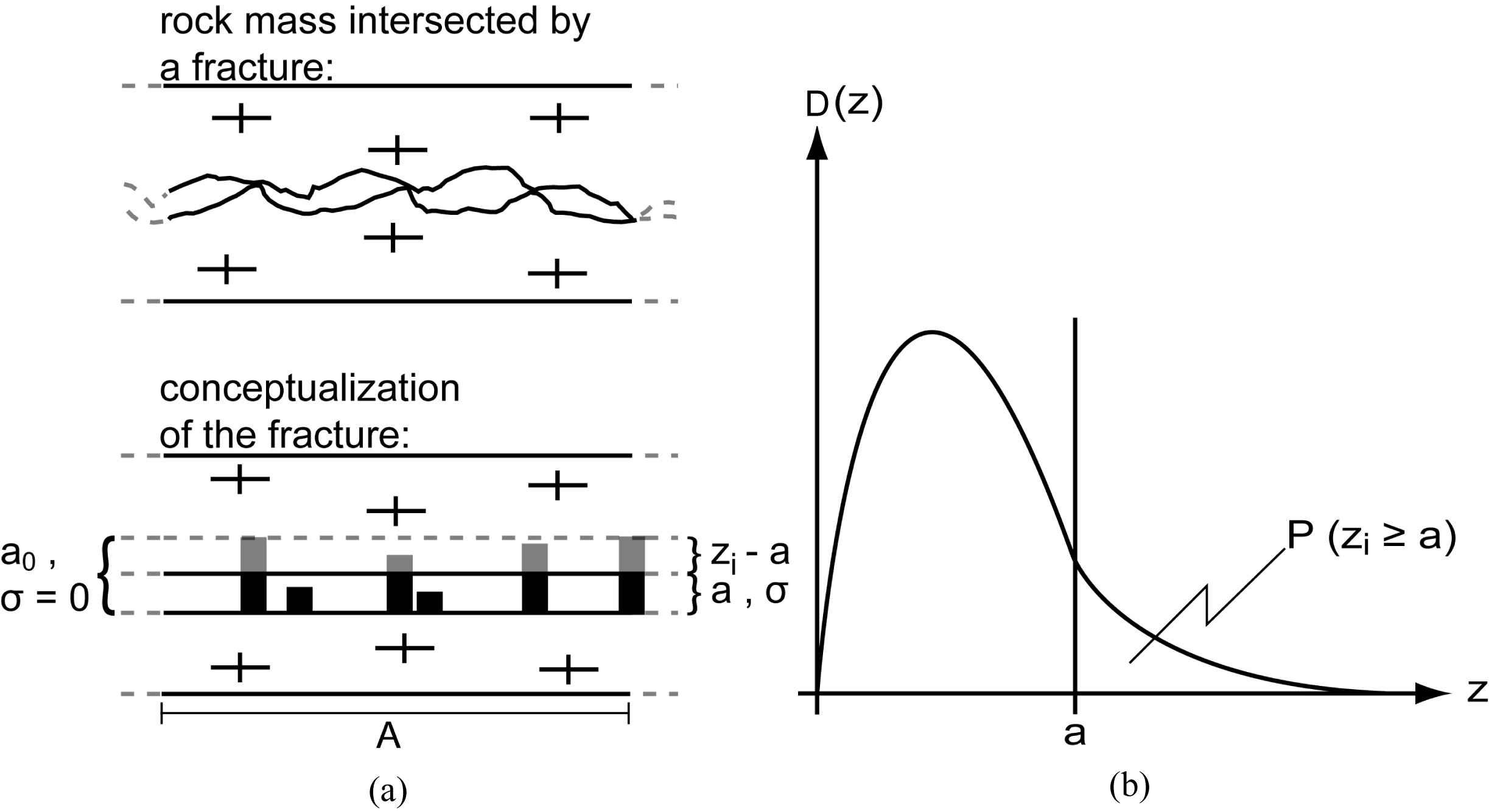}
  	\end{center}
  \caption{\small (a) Schematic illustration of a rock mass intersected by a fracture and its conceptualization with a set of asperities. The fracture under the normal stress $\sigma$ has the aperture $a$, the maximum fracture aperture $a_0$ is reached when there is no stress. (b) Continuous statistical distribution, $D(z)$, of the asperity length, $z$, and probability of contact.}
 \label{fig:Model}
\end{figure*}
Under a given normal stress, the asperities return an equilibrium equivalent stress $\sigma$ defining a specific aperture $a$ (Figure \ref{fig:Model}a). This normal stress results from the integration of all stresses exerted by individual asperities compressed to various degrees. For a given aperture, the probability that an asperity is in contact with both fracture faces $P(z_i \geq a)$ corresponds to the ratio between the number of compressed asperities and the total number of asperities:
	\begin{equation}
		P(z_i \geq a) = \int^{\infty}_{a} D(z) dz = \frac{N_c}{N_t}
		\label{eq:CompressionProbability}
	\end{equation}
where $D(z)$ is the statistical distribution of the asperity lengths (Figure \ref{fig:Model}b), $N_c$ is the number of compressed asperities and $N_t$ is the total number of asperities. In Eq. (\ref{eq:CompressionProbability}) the infinite upper bound of the integral can be replaced by the maximum fracture aperture $a_0$, which also represents the original length of the longest asperities. Note that the integral of $D(z)$ must be equal to unity. Glover et al. \cite{Glover1998} note that $D(z)$ is frequently assumed of Gaussian type. In this paper, a number of simple typical distributions are considered, as well as the more realistic Weibull distribution.

Assuming average values for asperity elastic modulus and section, and associating \ref{eq:CompressionProbability} and \ref{eq:HookeLaw}, the equilibrium normal stress for a fracture with an aperture $a$ is obtained by weighting each asperity contribution by its probability density. Integrating over all active asperities yields:
	\begin{equation}
		\sigma = \frac{F}{A} = \frac{N_t}{A} E s \int^{a_{0}}_{a} \frac{(z-a)}{z} D(z) dz
		\label{eq:ReturnedStress}
	\end{equation}
where $F$ is the force exerted by the compressed asperities, $A$ is the fracture surface area, $E$ is the elastic modulus of the fractured rock and $s$ is an average asperity section. $N_t/A=\eta$ is the asperity areal density. Eq. (\ref{eq:ReturnedStress}) respects the same conditions as Eq. (\ref{eq:HookeLaw}). Firstly, the maximum fracture aperture $a_0$ is reached at $\sigma=0$ ($a=a_0$, no compression). Secondly, total fracture compression ($a=0$) occurs when $\sigma=\eta E s=\sigma_0$, where $\sigma_0$ is the fracture closure normal stress.
\section*{\large{Model adjustment for different statistical distributions}}
\begin{figure*}[!ht]
  \begin{center}
  	\includegraphics[width=0.75\textwidth]{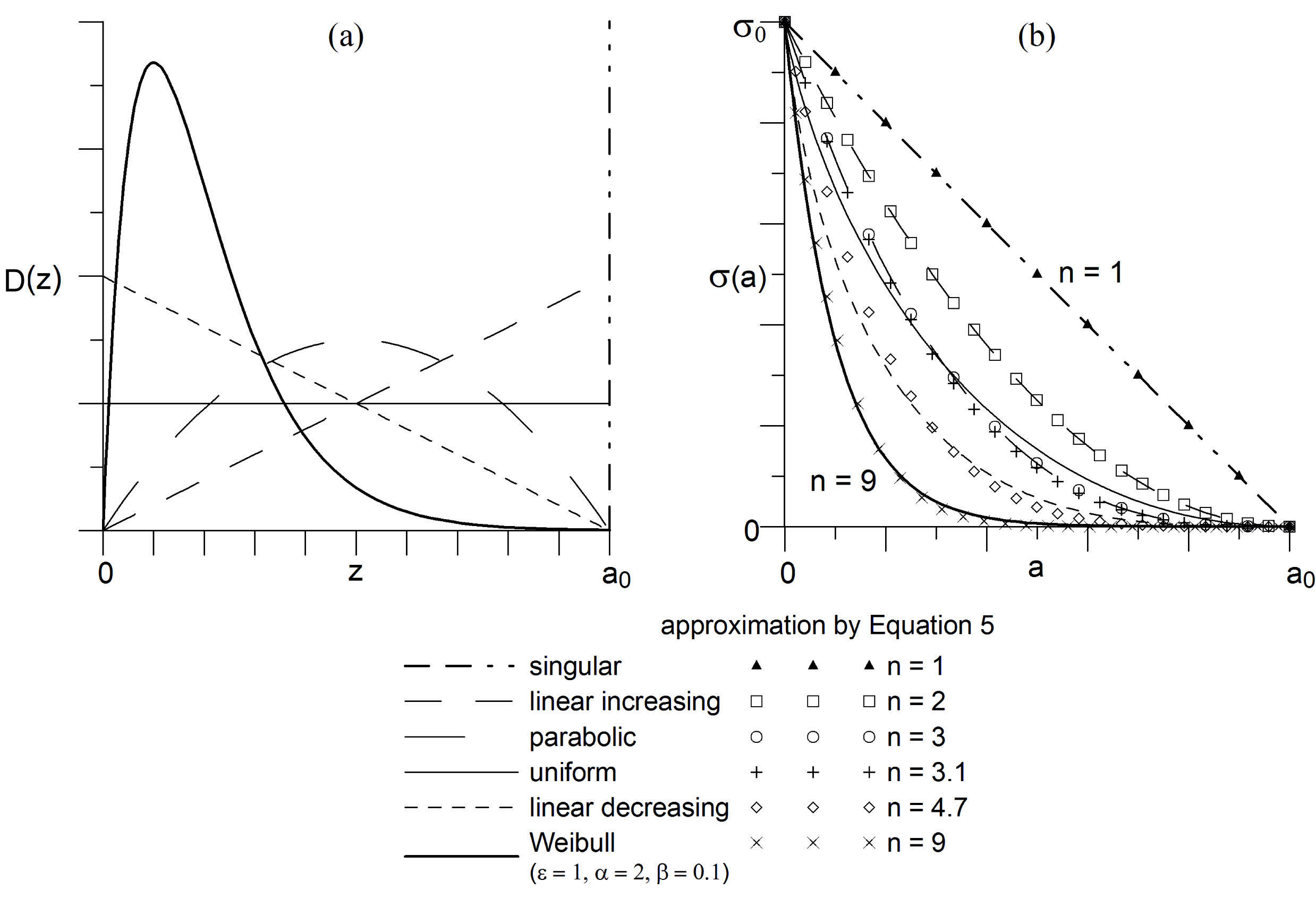}
  	\end{center}
  \caption{\small (a) Possible statistical continuous distributions of asperity length in a fracture, and (b) corresponding stress/aperture solutions, with their approximation by Eq. (\ref{eq:approximation}).}
 \label{fig:DistrApprox}
\end{figure*}

Different $\sigma(a)$ models are obtained depending on the statistical distribution $D(z)$. For example, for the uniform distribution (Figure \ref{fig:DistrApprox}a) $D(z)=1/a_0$ , Eq. (\ref{eq:ReturnedStress}) becomes:
\begin{eqnarray}
	\sigma & = & \sigma_{0} \int^{a_{0}}_{a} \frac{(z-a)}{z} \frac{1}{a_{0}} dz \\
	& = & \sigma_{0} \left[1-\frac{a}{a_{0}}+\frac{a}{a_{0}}\ln\left(\frac{a}{a_{0}}\right)\right] \nonumber
	\label{eq:UniformDistr}
\end{eqnarray}	
Several constitutive models are found proceeding in the same way for different types of distribution $D(z)$ (Table 1). After integration the fracture aperture $a$ is directly related to normal stress $\sigma$. As mentioned in the introduction, the model must be simple from a numerical point of view, hence the need to reformulate the specific models presented in Table 1 in a generic equation of the form:
\begin{equation}
	\sigma = \sigma_{0} \left(1-\frac{a}{a_{0}}\right)^n , n \geq 1
	\label{eq:approximation}
\end{equation}
\begin{table*}
\caption{\small Example of aperture-stress models for different statistical distributions of asperity length.}
\begin{center}
\begin{tabular}{l|l|l}
 & \small statistical distribution & \small model \\
\hline
\small singular & $D(z) = \delta(z - a_0)$ & $\sigma = \sigma_{0}\left(1-\frac{a}{a_{0}}\right)$ \\
 & & \\
\small uniform & $D(z) = \frac{1}{a_{0}}$ & $\sigma = \sigma_{0} \left[1-\frac{a}{a_{0}}+\frac{a}{a_{0}}\ln\left(\frac{a}{a_{0}}\right)\right]$ \\
 & & \\
\small linear increasing & $D(z) = \frac{2z}{a_{0}^2}$ & $\sigma = \sigma_{0}\left(1-\frac{a}{a_{0}}\right)^2$ \\
 & & \\
\small linear decreasing & $D(z) = \frac{2}{a_{0}} \left(1-\frac{z}{a_{0}}\right)$ & $\sigma = \sigma_{0}\left[1-\left(\frac{a}{a_{0}}\right)^2+2\frac{a}{a_{0}}\ln\left(\frac{a}{a_{0}}\right)\right]$ \\
 & & \\
\small parabolic & $D(z) = \frac{6}{a_{0}^2}z\left(1-\frac{z}{a_{0}}\right)$ & $\sigma = \sigma_{0} \left(1-\frac{a}{a_{0}}\right)^3$ \\
 & & \\
\small Weibull & $D(z) = \frac{\epsilon}{\beta a_{0}} \left(\frac{z}{\beta a_{0}}\right)^{(\alpha-1)} \times $ & $\sigma = \frac{\sigma_{0}}{C}\left[e^{(-10\frac{a}{a_{0}})}+C-1+\frac{a}{a_{0}}e^{(-10)}\right]$ \\
\small distribution & $e^{\left[-\left(\frac{z}{\beta a_{0}}\right)^{\epsilon}\right]} \frac{1}{C}$ & \scriptsize{for $\epsilon = 1$, $\alpha = 2$ and $\beta = 0.1$ ($C = 0.9995$)}\\
 & \scriptsize{C: normalization constant so that} & \\
 & \scriptsize{$\int^{\infty}_{0}D(z)dz = 1$} & \\
\end{tabular}
\label{tab:Models}
\end{center}
\end{table*}
Depending on the value of coefficient $n$, Eq. (\ref{eq:approximation}) provides exact stress-dependent apertures for the non-logarithmic functions of Table 1 and good approximations for the logarithmic ones (Figure 2b). The symbol n stands for the coefficient of asperities length statistical distribution.
Statistical distributions characterized by many large asperities, such as singular and linear increasing, get low coefficients $n$, 1 and 2, respectively. On the contrary, distributions with many small asperities (linear decreasing and Weibull) are correctly approximated with relatively high coefficients $n$ (4.7 and 9).
Inversely, the distribution $D(z)$ can be found for a given stress/aperture function $\sigma(a)$. Differentiating Eq. (3) twice with respect to $a$ yields:
\begin{equation}
	\frac{\partial^2 \sigma(a)}{\partial a^2} = \frac{\sigma_{0}}{a} D(a)
	\label{eq:equality}
\end{equation}
For the general model assumed in Eq. (\ref{eq:approximation}), this yields:
\begin{equation}
	\frac{\partial^2 \sigma(a)}{\partial a^2} = \sigma_{0} \frac{n(n-1)}{a_0^2}\left(1-\frac{a}{a_0}\right)^{n-2}
	\label{eq:equality_for_model}
\end{equation}
and the distribution $D(z)$ is obtained by equating Eqs. (\ref{eq:equality}) and (\ref{eq:equality_for_model}):
\begin{equation}
	D(z) = n(n-1) \frac{z}{a_0^2}\left(1-\frac{z}{a_0}\right)^{n-2} , n \geq 1
	\label{eq:ditribution_for_model}
\end{equation}
\section*{\large{Relation with hydrogeological parameters}}
Because of the saturated flow conditions considered in this work (i.e. fractures are completely filled by water exerting the pressure p), the normal effective stress $\sigma'$ is taken into account instead of the normal total stress $\sigma$. In the case of lithostatic stress conditions $\sigma_z=\rho_r g Z$ and in the absence of shear stresses, the resulting effective stress $\sigma'$ acting perpendicularly at a depth $Z$ on a given fracture plane is obtained by:
\begin{eqnarray}
  \sigma' & = & \mathrm{\mbox{\boldmath${\sigma}$}} \mathbf{n} \cdot \mathbf{n} \hspace{0.1cm}  - p\hspace{0.1cm}, \hspace{0.5cm} \mathrm{\mbox{\boldmath${\sigma}$}} = \left[
	\begin{array}{c c c}
	 	 \sigma_z \lambda & 0 & 0 \\
	 	0 & \sigma_z \lambda & 0 \\
		0 & 0 & \sigma_z \\
	\end{array} \right] \nonumber\\
	        & = & \rho_r g Z (\lambda n_x^2 + \lambda n_y^2 + n_z^2) - \alpha\rho_w g h \label{eq:Terzaghi} \\
		\nonumber
\end{eqnarray}
where $\rho_r$ is the rock mass density, $g$ is the gravitational acceleration, $n_x$,$n_y$,$n_z$ are the components of the unit vector $n$ normal to the fracture plane, $\rho_w$ is the water density, $h$ is the pressure head, and $\alpha$ is the Biot-Willis coefficient. The $\lambda$ coefficient is the ratio of horizontal to vertical stress.

Expressing Eq. (\ref{eq:approximation}) for the aperture $a$:
\begin{equation}
	a = a_0 \left[1 - \left(\frac{\sigma'}{\sigma'_0}\right)^{\frac{1}{n}}\right]
	\label{eq:ApertureModel}
\end{equation}
and assuming the validity of the cubic law in the fractured rock, the stress-dependent permeability is:
\begin{equation}
	k = \frac{f a^3}{12} = \frac{f a_0^3 \left[1 - \left(\frac{\sigma'}{\sigma'_0}\right)^{\frac{1}{n}}\right]^3}{12}
\label{eq:CubicLaw}	
\end{equation}
yielding the hydraulic conductivity parallel to fracture plane:
\begin{equation}
	K = K_0 \left[1 - \left(\frac{\sigma'}{\sigma'_0}\right)^{\frac{1}{n}}\right]^3
	\label{eq:Kmodel}
\end{equation}
where:
\begin{equation}
	K_0 = \frac{\rho_w g}{\mu_w} \frac{f a_0^3}{12}
	\label{eq:MaxHydrConductivity}
\end{equation}
with a maximum $K_0$ for $\sigma'=0$. The symbol $f= N_f/d$ is the frequency of the fracture family, namely the number of fractures $N_f$ counted over a distance $d$, and $\mu_w$ is water viscosity. Note that Eq. (\ref{eq:Kmodel}) is very similar to the constitutive models proposed by \cite{Lombardi1992, Li2001}. The same model function was found by \cite{Gangi1978} via a different approach. Eq. (\ref{eq:Kmodel}) can be used to compute the equivalent macroscopic hydraulic conductivity tensor of a rock mass intersected by $m$ fracture families using the tensor summation:
\begin{equation}
	\mathbf{K} = \sum^{m}_{i=1} K_{0_i} \left[1 - \left(\frac{\sigma'_i}{\sigma'_{0_i}}\right)^{\frac{1}{n_i}}\right]^3 \left(\mathbf{I} - \mathbf{n}_i \otimes \mathbf{n}_i\right)	
	\label{eq:TensorKmodel}
\end{equation}
for each fracture family $i$, $K_{0_i}$ is the maximum parallel hydraulic conductivity, $\sigma'_i$ is the normal effective stress, $\sigma'_{0_i}$ is the fracture closure normal stress, $n_i$ relates to the asperity distribution, $I$ is the identity matrix, $n_i$ is the unit vector normal to the fracture family $i$, and $\otimes$ denotes a tensor product.

If the contribution of the rock matrix is neglected, the porosity $\phi$ of the fractured rock mass is:
\begin{equation}
	\phi = \sum^{m}_{i=1} f_i a_i
	\label{eq:Porosity}
\end{equation}
Introducing Eq. (\ref{eq:ApertureModel}) into Eq. (\ref{eq:Porosity}), a stress-dependent porosity is obtained:
\begin{equation}
	\phi = \sum^{m}_{i=1} \phi_{0_i} \left[1 - \left(\frac{\sigma'_i}{\sigma'_{0_i}}\right)^{\frac{1}{n_i}}\right]
	\label{eq:PhiModel}
\end{equation}
Always neglecting the contribution of the rock matrix, Eq. (\ref{eq:PhiModel}) can be introduced in the definition of the specific storage coefficient, $S_s$:
\begin{eqnarray}
	S_s & = & \frac{\rho_w g \phi}{E_w} \nonumber \\
	S_s & = & \sum^{m}_{i=1} S_{s_{0_i}} \left[1 - \left(\frac{\sigma'_i}{\sigma'_{0_i}}\right)^{\frac{1}{n_i}}\right] ; \label{eq:SsModel} \\
	S_{s_{0_i}} & = & \frac{\rho_w g \phi_{0_i}}{E_w} \nonumber \\
	\nonumber
\end{eqnarray}
where $\phi_0$ and $S_{s_0}$ are the maximum fracture porosity and maximum specific storage coefficient, respectively. Eq. (\ref{eq:PhiModel}) can be used to determine, for each fracture family $i$, the vertical variation in fracture porosity, $\Delta\phi$, due to a change in effective stress, between an initial and a successive hydrogeological state at elevation $z$:	 
\begin{eqnarray}
\Delta\phi(z) & = & \sum^{m}_{i=1} \left(\phi_{h_i} - \phi_{h_{s_i}}\right) n_{z_i} \nonumber \\
              & = & \sum^{m}_{i=1} \phi_{0_i} \left[ \left(\frac{\sigma'_{h_{s_i}}}{\sigma'_{0_i}}\right)^{\frac{1}{n_i}} - \left(\frac{\sigma'_{h_i}}{\sigma'_{0_i}}\right)^{\frac{1}{n_i}} \right] n_{z_i} \label{eq:DeltaPorosity} \nonumber \\
\end{eqnarray}	
where the symbols $\phi_h$ and $\phi_{h_s}$ stand for fracture porosity at an initial and at a successive pressure head state. The multiplication with the component $n_z$ of the unit normal vector $n$ is used to obtain the vertical change in fracture porosity. Integrating all the porosity changes in the vertical direction, from the bottom of the aquifer $z_b$ to the top $z_t$, results in the local settlement:
\begin{equation}
T(x,y) = \int^{z_t}_{z_b} \Delta\phi(z)\hspace{0.1cm}dz
	\label{eq:Consolidation}
\end{equation}
Eq. (\ref{eq:Consolidation}) provides: (1) where $T>0$ is aquifer vertical consolidation under increasing effective stress and (2) where $T<0$ is aquifer vertical expansion under decreasing effective stress.
Finally, considering both the hydraulic conductivity and the specific storage coefficient as functions of effective stress results in the non-linear groundwater flow equation:
\begin{equation}
  S_s(\sigma') \frac{\partial H}{\partial t} = \nabla\cdot\left(\mathbf{K}(\sigma') \nabla H\right) \hspace{0.3cm} ; \hspace{0.3cm} H = h + z
	\label{TransientFlowEquation}
\end{equation}
where $H$ is the hydraulic head, $\mathbf{K}(\sigma')$ is the hydraulic conductivity tensor as expressed in Eq. (\ref{eq:TensorKmodel}), $S_s(\sigma')$ is the specific storage coefficient as expressed in Eq. (\ref{eq:SsModel}), and $t$ is time. The symbols $h$ and $z$ stand for the relative pressure and elevation head, respectively.
\section*{\large{Comparison between simulated, experimental and field measured permeabilities}}
\begin{figure*}[!ht]
  \begin{center}
  	\includegraphics[width=0.75\textwidth]{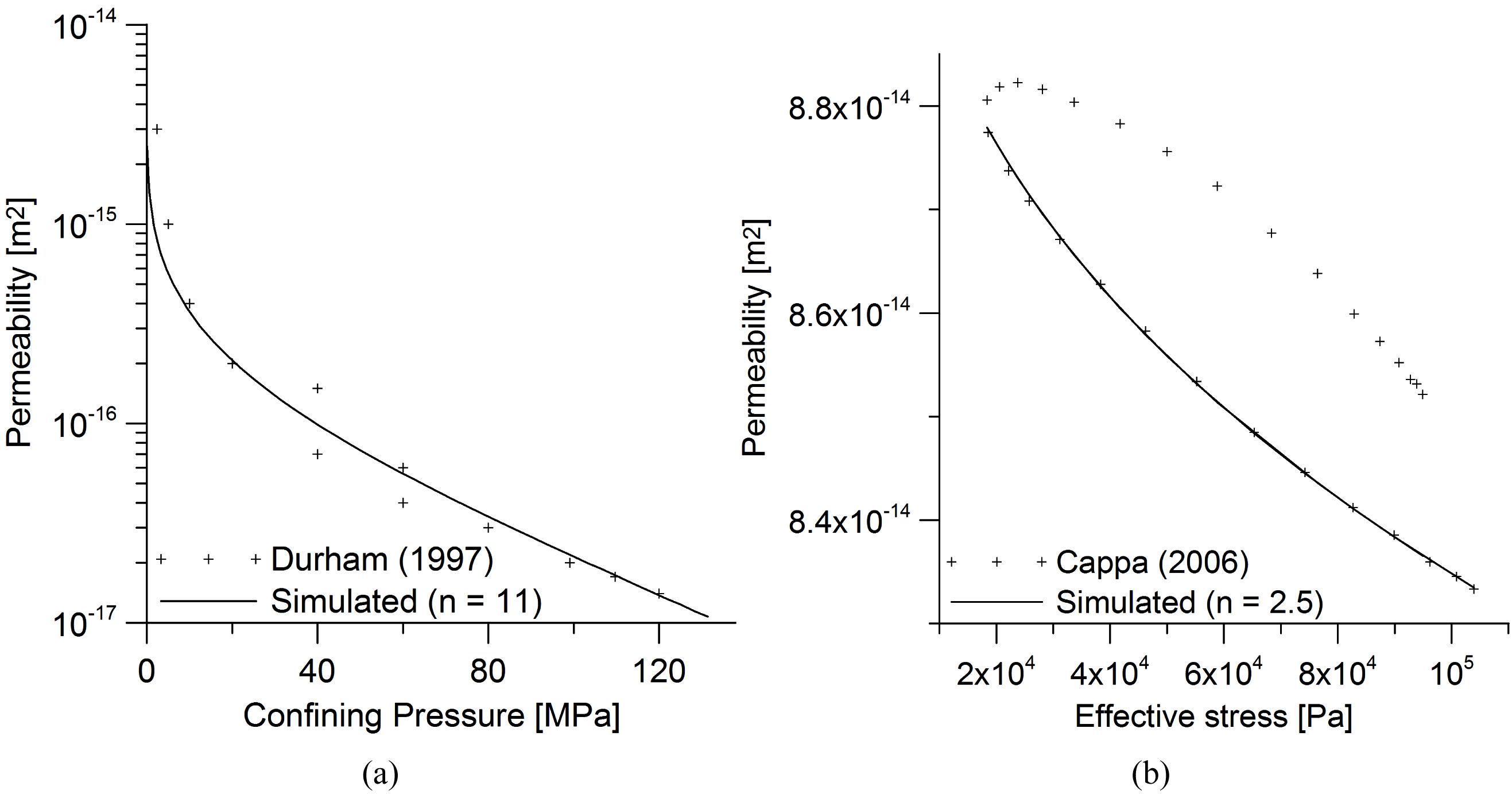}
  \end{center}
  \caption{\small Verification of Eq. (\ref{eq:Kmodel}) model by comparison with stress-dependent fractured rock permeabilities of (a) Durham \cite{Durham1997} and (b) Cappa \cite{Cappa2006}.}
 \label{fig:DurhamCappa}
\end{figure*}

Eq. (\ref{eq:Kmodel}) is verified by comparison with stress-dependent permeabilities from Durham \cite{Durham1997} and Cappa \cite{Cappa2006}.
Laboratory tests carried out by Durham \cite{Durham1997} showed the behavior of the permeability of a fracture sample, taken at approximately 3.8 km depth, when subjected to an increasing confining pressure (stress). Simulated permeabilities correspond well to those measured by Durham \cite{Durham1997}, especially for high stresses (Figure \ref{fig:DurhamCappa}a, Appendix II).
With experiments at shallow conditions, Cappa \cite{Cappa2006} investigated the pressure-dependent increase and decrease of fracture aperture. Results showed that fracture aperture is subjected to hysteresis process. Eq. (\ref{eq:Kmodel}) is used to fit the field data of Cappa \cite{Cappa2006} (Figure \ref{fig:DurhamCappa}b), which for that example were transformed from aperture and water pressure to permeability and normal effective stress. Fitted parameters are given in Appendix II. Also for this example, the model provides a good comparison between simulated and measured data. However, only the rising branch of the hysteresis curve, which corresponds to an increasing water pressure and a decreasing effective normal stress, is correctly simulated. Eq. (\ref{eq:Kmodel}) can not reproduce a hysteresis, because the model describes only the elastic part of deformation, and therefore simulated permeabilities will be the same for rising or falling effective normal stresses. In Figure \ref{fig:DurhamCappa}b the hysteresis occurs because the tested rock does not exactly follow Hooke's law. However the variation of permeability is so low, that this phenomenon may be neglected at regional scale.
Note that, numerical values of the coefficient of asperities length statistical distribution, $n$, and the fracture closure effective stress, $\sigma'_0$, can be obtained by calibration of Eq. (\ref{eq:Kmodel}) on measured stress-dependent permeability data, and applied for large scale analysis.
\section*{\large{Illustrative examples}}
\subsection*{Steady state}
The preceding equations were implemented in the multi-purpose Groundwater finite element software \cite{Cornaton2007}, in order to illustrate: (1) how simulation results vary, if the effect of effective stress on hydrogeological parameters is taken into account; (2) the regional effects of a deep tunnel.
At steady state and with stress-dependent hydraulic conductivity the flow equation is:
\begin{equation}
	\nabla\cdot\left(\mathbf{K}(\sigma') \nabla H\right) = 0 \hspace{0.3cm} ; \hspace{0.3cm} H = h + z
	\label{eq:FlowEquation}
\end{equation}
The virtual model domain is a 2D vertical cross section representing an Alpine hydrogeological system composed of three geological formations, completely saturated with water, with different hydraulic properties (Figure \ref{fig:Sim_Domain}, Appendix II). Note that, in Appendix II are shown numerical values used in simulations, these are based on field investigations of the Emosson fractured rock mass (Switzerland), using the method described by Király \cite{Kiraly1969b}. In appendix II, $K_{max}$ and $K_{min}$ are the eigenvalues of the hydraulic conductivity tensor, and $\theta$ is the angle between the horizontal plane and the direction of $K_{max}$. The stress field is defined by vertical stresses $\sigma(z)$ set equal to the lithostatic pressure (the weight of overlying rocks above elevation z):
\begin{equation}
	\sigma(z) = g \int^{z_t}_{z} \rho_r(u)\hspace{0.1cm}du
	\label{LithostaticStress}
\end{equation}
\begin{figure*}[!ht]
  \begin{center}
  	\includegraphics[width=0.65\textwidth]{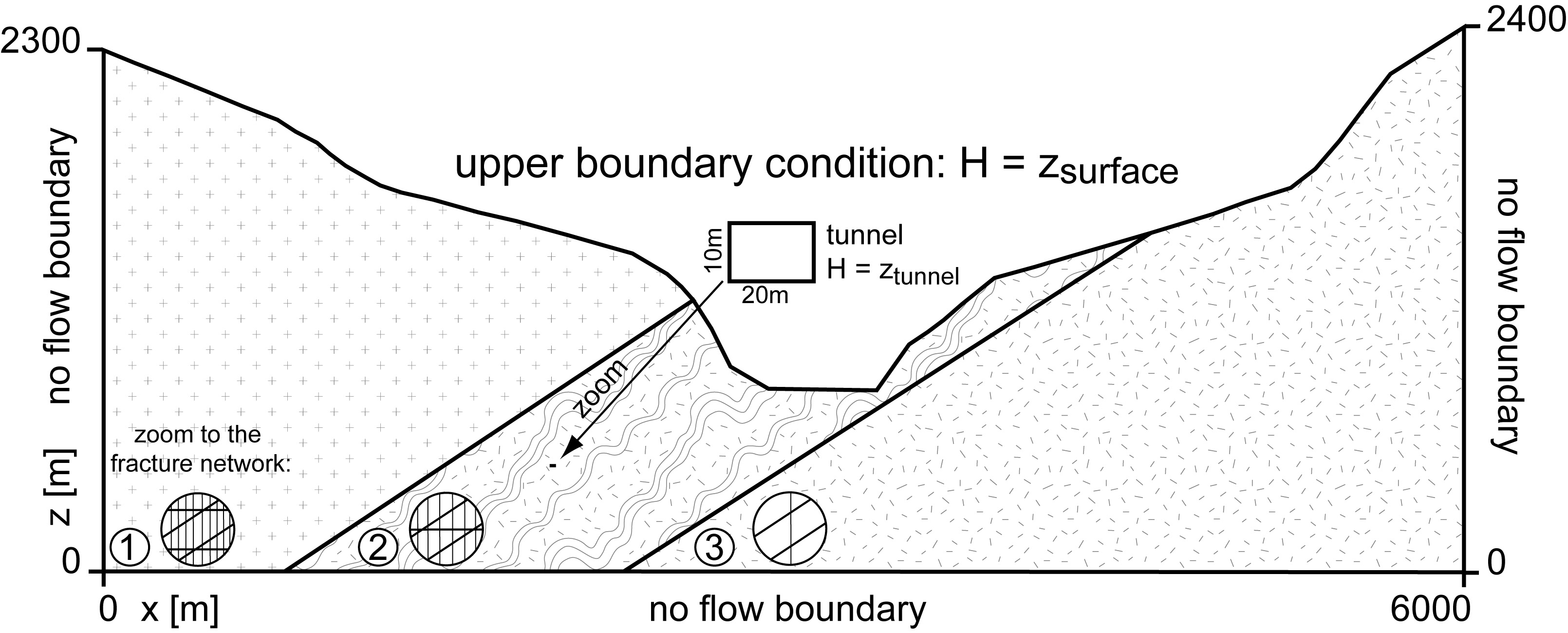}
  	\end{center}
  \caption{\small Model domain and boundary conditions; the hydrogeological system is composed of three rocks, each one exhibiting a different fracture network. Rock 2 is the most permeable, while rock 3 is the least. Three simulations are run for each statistical distribution of the asperities length: one without tunnel (natural system), one with tunnel, and finally one that simulates the aquifer consolidation.}
 \label{fig:Sim_Domain}
\end{figure*}
In the present study horizontal stresses $\sigma_x$ are 1.5 times stronger than the vertical stresses. This applies well to orogenic belts or areas that have been glaciated, such as the Alps (see \cite{Mayeur1999}). Before tunnel construction, a steady state flow is assumed from the highest points (crests) to the lowest points (valleys), by specifying boundary conditions at the domain surface as atmospheric pressure ($H=z$), and at other limits as no-flow conditions. Then, a tunnel is constructed. A constant atmospheric pressure is specified in the tunnel indicating that it behaves as a draining structure and consequently increases the effective stress which causes aquifer consolidation. Several simulations are computed to compare the present approach with the classical one neglecting the dependence of permeability on effective stress, and to study the influence of the coefficient $n$ on discharge rates, pressure and consolidation distributions. Consolidation is computed between the initial state (without tunnel) and the disturbed state with perturbation caused by the tunnel (Figure \ref{fig:Sim_Fields}).
This illustrative model is directly inspired by real cases of fractured aquifer consolidation caused by tunnels excavation (see \cite{Lombardi1988}, and \cite{Zangerl2003}).
\subsection*{Results and discussions}
\begin{figure*}[!ht]
  \begin{center}
  	\includegraphics[width=0.75\textwidth]{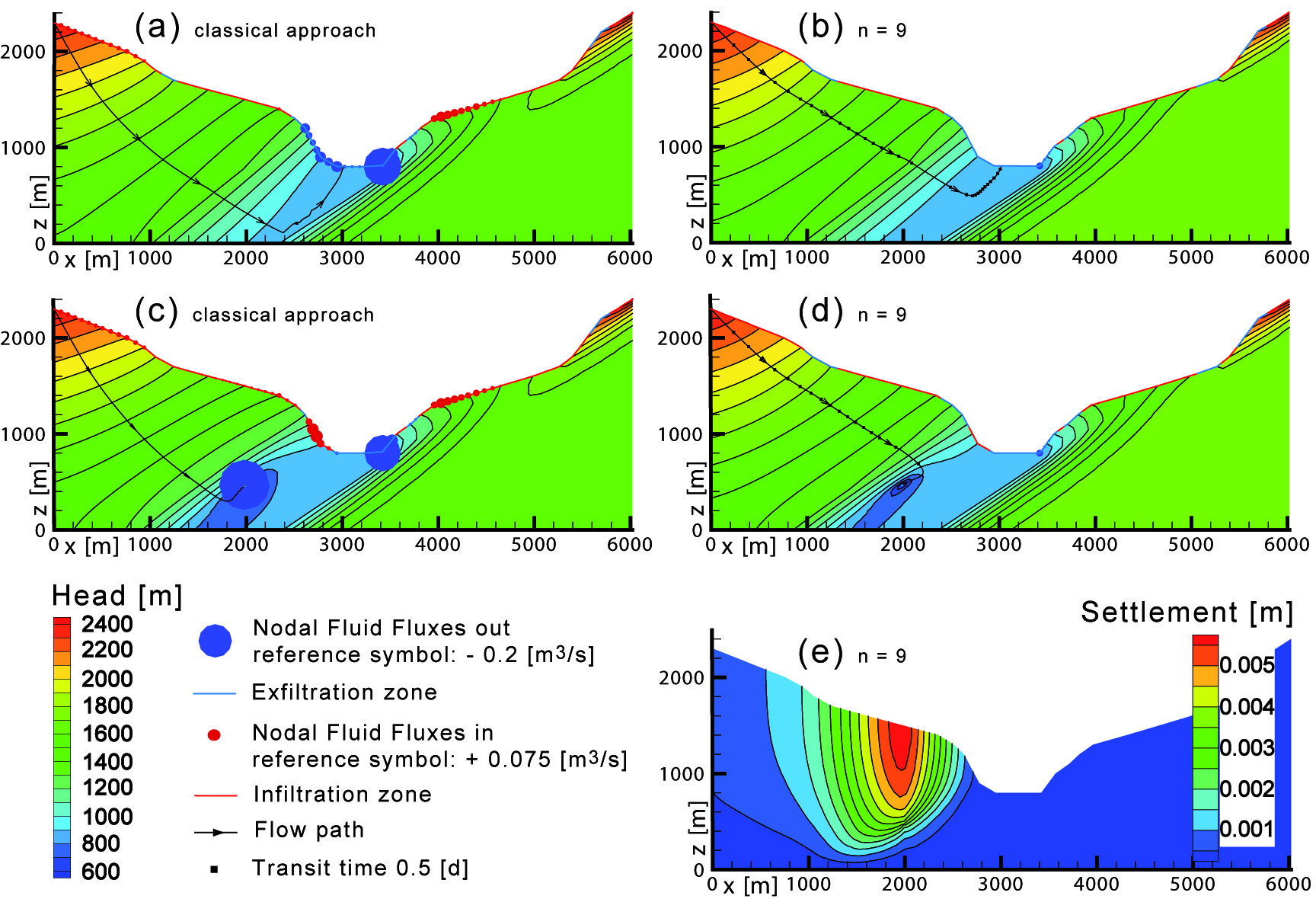}
  \end{center}
  \caption{\small Hydraulic head, flow paths and infiltration/exfiltration fields for the classical approach, and for the Weibull distribution at initial conditions (a, b) and after the tunnel introduction (c, d). Note that, in (b) and (d) fluid fluxes are so much lower than in (a) and (c), that they are almost invisible. (e) Aquifer consolidation caused by the increase in effective stress following the tunnel construction (Weibull distribution).}
 \label{fig:Sim_Fields}
\end{figure*}

\begin{figure*}[!ht]
  \begin{center}
  	\includegraphics[width=0.75\textwidth]{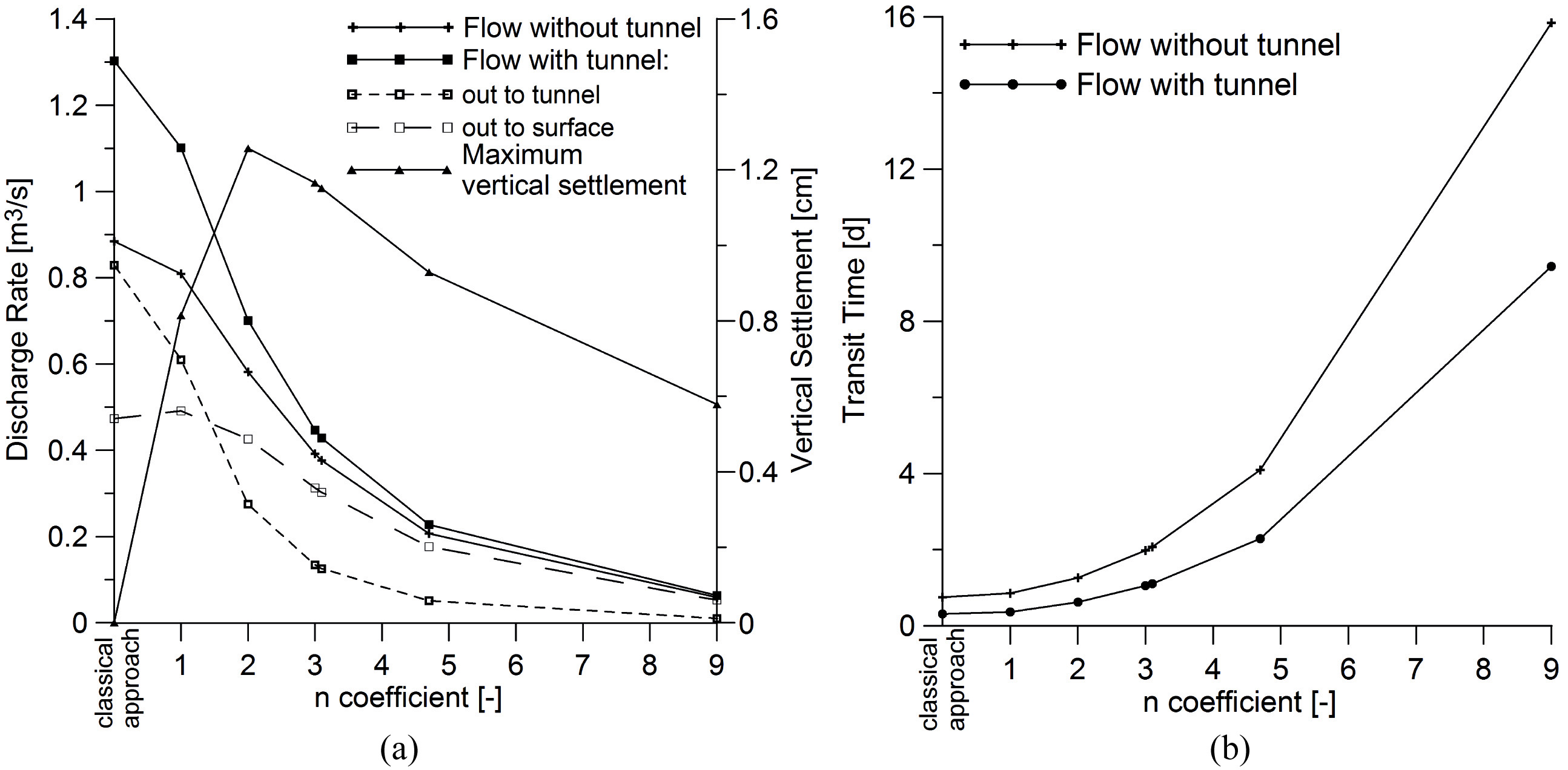}
  \end{center}
  \caption{\small (a) Discharge rates, vertical settlement and (b) transit time as a function of coefficient n, and comparison with the classical approach for steady state flow before and after the tunnel construction. For transit time the particle is released at coordinates $x=0$ and $z=2300$, and exits: (1) at the bottom of the valley (without tunnel); (2) at the tunnel (particle tracks are shown in Figure \ref{fig:Sim_Fields} for the classical approach and for the Weibull distribution).}
 \label{fig:Sim_graphics}
\end{figure*}

Results show that the introduction of stress-dependent permeabilities in Darcy's law leads to lower discharge rates, relative to the classical approach that only considers constant permeability, especially for high values of coefficient $n$ (Figure 6a). This reduction in discharge rates is directly related to the variations of the hydraulic conductivity tensor in both strength and principal directions according to effective stresses. This spatial variation of hydraulic conductivity tensors also impacts the distribution and the magnitude of hydraulic heads, flow paths, flow velocities and transit times (Figures \ref{fig:Sim_Fields} and \ref{fig:Sim_graphics}b). Overall permeabilities decrease in the deeper areas of the domain, while they tend toward the maximum near the surface. As previously mentioned, a high value of the coefficient $n$ indicates a predominance of relatively small asperities. In such a case, the drop in permeability will be significant, because there are only a few large asperities to oppose the increase in normal effective stress.
Compared to the classical approach, the impact of the tunnel on the system appears weaker when considering stress-dependent permeabilities.
For the proposed method, the highest consolidation occurs in systems with an intermediate $n$ value ($1 < n < 5$), because they are the most sensitive to pressure change with the largest porosity variation (Figure \ref{fig:Sim_graphics}a). Fractured systems featuring large asperities are less affected by the process, because the asperities stop the closure. Overall, the magnitude of aquifer consolidation is low because the proposed method computes only elastic reversible deformations obeying Hooke's law, and acting on fracture network porosity. Moreover, boundary conditions specified at the domain surface provide unlimited water inflows that dampen aquifer depressurization.
\subsection*{Transient state}
Eq. (\ref{TransientFlowEquation}) is used to solve the transient groundwater flow problem having the same model domain, boundary conditions at the tunnel and hydrological parameters. The initial hydraulic heads are taken from the steady state model without tunnel. On the domain surface two different types of boundary conditions are tested: (1) constant atmospheric pressure ($H=z$); (2) no-flow condition. This no-flow condition could represent an aquifer filled with connate pore waters and isolated from recharge zones, or a confined aquifer suddenly cut-off from its recharge zone.
\subsection*{Results and discussions}
\begin{figure*}[!ht]
  \begin{center}
  	\includegraphics[width=0.75\textwidth]{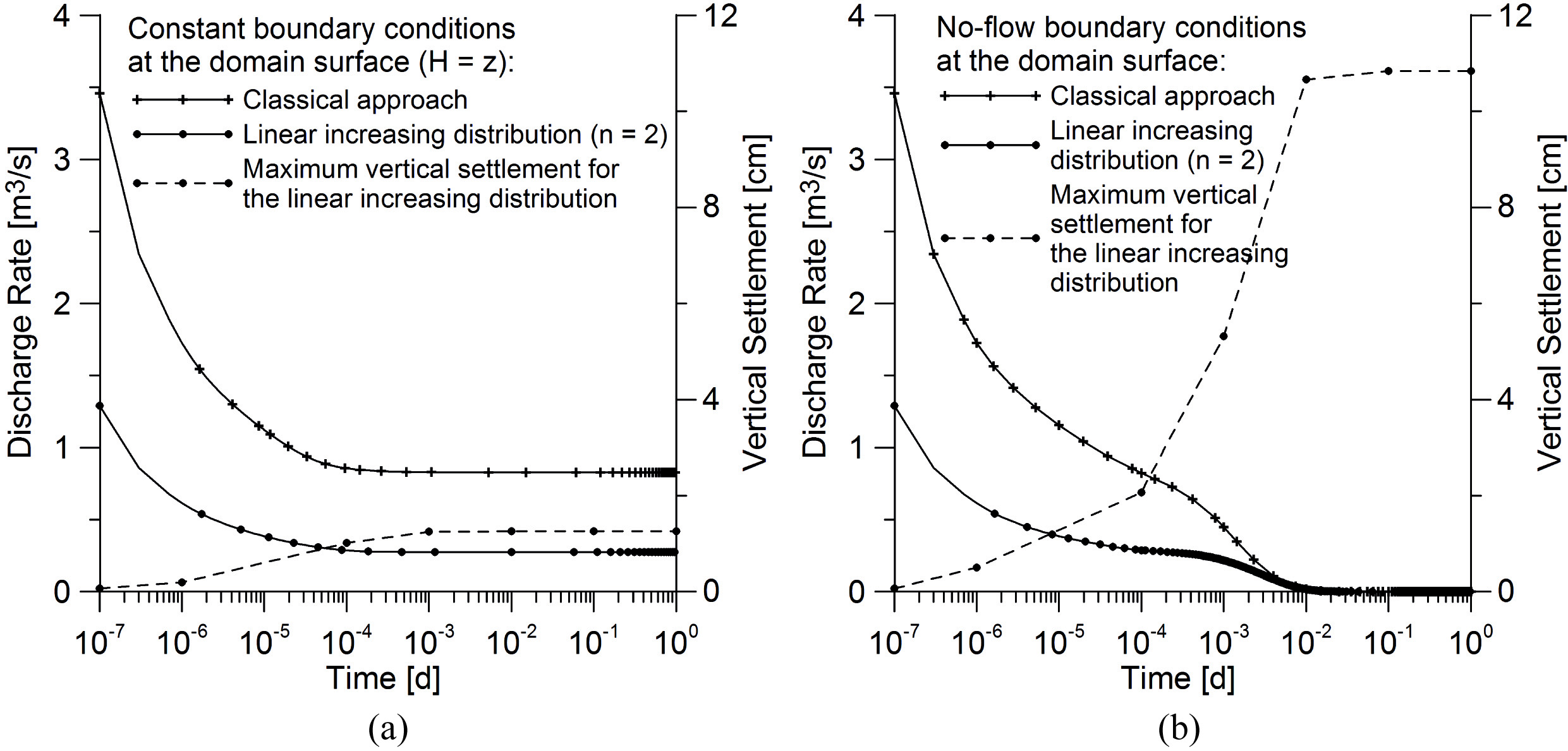}
  \end{center}
  \caption{\small Evolution of the discharge rate drained by the tunnel (for the classical approach and the linear increasing distribution, solid lines), and of the maximum vertical settlement (for the linear increasing distribution, dashed line) as a function of time for (a) constant hydraulic head at the domain surface; (b) no-flow condition at the domain surface.}
 \label{fig:T_D_S}
\end{figure*}

For the first case where temporally constant atmospheric pressure hydraulic heads are specified at the domain surface, the initial and final discharge rates as well as the vertical settlements match those simulated by the steady state models. In transient state, the tunnel causes a hydraulic depressurization of the rock mass followed by gradual aquifer consolidation (Figure \ref{fig:T_D_S}a).
For the second case with sudden no-flow condition at the domain surface, the tunnel drainage empties the system, which becomes hydrostatic. The recession curve of the water drained by the tunnel rapidly runs dry. In such a case, the magnitude of the aquifer consolidation increases because of the total depressurization of the system (Figure \ref{fig:T_D_S}b).
Overall, the transient state is relatively fast because there is no release of water from the rock matrix, assumed impervious.
\subsection*{Conclusions}
A model function relating fracture permeability to effective stress is derived from Hooke's law of elasticity and from the statistical distribution of asperity lengths. This model function is then implemented in the tensor form of Darcy's law, and its effects are assessed in simulations. Taking into account the sensitivity of permeabilities to depth and water pressure, this non-linear approach gains in phenomenology and is closer to physical reality, compared to classical approaches that neglect pressure-dependent permeability and porosity fields.
From a general point of view, numerical simulations of deep tunnels considering the decrease in permeability with increasing effective stress generate lower discharge rates. This observation goes in the line of safety in terms of problems due to the presence of water in the underground structure. In case of strong decrease in water pressures a non-negligible consolidation occurs, even when flow is assumed in the fractures only. This can produce foundation instabilities of structures located at the surface, especially in the case of differential consolidation due to aquifer heterogeneity, and in the presence of heavy structures such as dams (see \cite{Lombardi1988}).
The changes in fracture permeability, porosity and specific storage in response to changes in effective stresses depend on, among other factors, the statistical distribution of asperity lengths, indicating the relative ratio of large to small asperities. For example, in the case of increasing effective stresses, a fracture characterized by a high ratio of large to small asperities (small $n$), will have a lower change in permeability than a fracture with a small ratio (high $n$).
Overall, the limitation of the classical method is that it cannot compute aquifer consolidation, because no change in fracture porosity or permeability with pressure head variation is accounted for. On the contrary, with the proposed approach, a pressure head variation causing fracture porosity to change can be directly translated into aquifer consolidation (decreasing pressure) or expansion (increasing pressure).
\subsection*{Acknowledgments}
We gratefully acknowledge the constructive comments of Prof. Thomas J. Burbey and of the anonymous reviewers.
\appendix
\begin{figure*}[!ht]
\begin{center}
\scalebox{1.0}{
\begin{tabular}{l l}
\multicolumn{2}{l}{Appendix I: Notation} \\
\hline
$a$ [$m$] & fracture aperture (asperity length under compression) \\
$a_0$ [$m$] & maximum fracture aperture (original length of the longest asperities) \\
$A$ [$m^2$] & area \\
$E$ [$Pa$]  & fractured rock elastic modulus \\
$d$ [$m$] & distance \\
$D(z)$ [$-$] & statistical distribution of the asperities length \\
$f$ [$m^{-1}$] & frequency of the fracture family \\
$F$ [$N$] & force \\
$g$ [$m\hspace{0.1cm}s^{-2}$] & gravitational acceleration \\
$h$ [$m$] & pressure head \\
$H$ [$m$] & hydraulic head \\
$k$ [$m^2$] & intrinsic permeability \\
$k_0$ [$m^2$] & maximum intrinsic permeability \\
$K$ [$m\hspace{0.1cm}s^{-1}$] & hydraulic conductivity \\
$K_0$ [$m\hspace{0.1cm}s^{-1}$] & maximum hydraulic conductivity \\
$m$ [$-$] & number of fracture families \\
$n$ [$-$] & coefficient of asperities length statistical distribution \\
$n_x, n_y, n_z$ [$-$] & components of the unit normal vector \\
$N_c$ [$-$] & number of compressed asperities \\
$N_f$ [$-$] & number of fractures \\
$N_t$ [$-$] & total number of asperities \\
$p$ [$Pa$] & water pressure \\
$s$ [$m^2$] & average asperity section \\
$S_s$ [$m^{-1}$] & specific storage coefficient \\
$S_{s_0}$ [$m^{-1}$] & maximum specific storage coefficient \\
$t$ [$s$] & time \\
$T$ [$m$] & ground settlement or expansion\\
$z$ [$m$] & 1: asperity's original length, 2: elevation head \\
$Z$ [$m$] & depth \\
$\alpha$ [$-$] & Biot-Willis coefficient \\
$\eta$ [$m^{-2}$] & asperity areal density \\
$\lambda$ [$-$] & ratio of horizontal to vertical stress \\
$\mu_w$ [$kg\hspace{0.1cm}m^{-1}\hspace{0.1cm}s^{-1}$] & water viscosity \\
$\rho_r$ [$kg\hspace{0.1cm}m^{-3}$] & rock mass density \\
$\rho_w$ [$kg\hspace{0.1cm}m^{-3}$] & water density \\
$\sigma$ [$Pa$] & normal stress \\
$\sigma'$ [$Pa$] & normal effective stress \\
$\sigma_{0}$ [$Pa$] & fracture closure stress \\
$\sigma'_{0}$ [$Pa$] & fracture closure effective stress \\
$\phi$ [$-$] & porosity \\
$\phi_0$ [$-$] & maximum porosity \\
$\Delta\phi$ [$-$] & porosity variation \\
\end{tabular}
}
\end{center}
\end{figure*}
\begin{figure*}[!ht]
\begin{center}
\scalebox{1.0}{
\begin{tabular}{|l|l|l|l l l|}
\multicolumn{6}{l}{Appendix II: Used values} \\
\hline
 & Fig. 3a & Fig. 3b & Illustration & Simulation & \\
 & & & Rock 1 & Rock 2 & Rock 3 \\
\hline
$m$ [$-$] & 1 & 1 & 3 & 3 & 2 \\
$a_{0_1}$ [$mm$]   & - & 0.1 & 0.5  & 0.5 & 0.5 \\
$a_{0_2}$ [$mm$]   & & & 1.2 & 1.92 & 1.2 \\
$a_{0_3}$ [$mm$]   & & & 1.0 & 1.63 &     \\
$f_1$ [$m^{-1}$] & - & 1 & 5.44 & 3.27 & 0.1 \\
$f_2$ [$m^{-1}$] & & & 0.71 & 0.71 & 0.5 \\
$f_3$ [$m^{-1}$] & & & 1.00 & 0.01 & \\
$K_{max}$ [$m\hspace{0.1cm}s^{-1}$] & & & $1.56\cdot10^{-3}$ & $3.77\cdot10^{-3}$ & $6.31\cdot10^{-4}$ \\
$K_{min}$ [$m\hspace{0.1cm}s^{-1}$] & & & $5.47\cdot10^{-4}$ & $2.1\cdot10^{-4}$ & $6.1\cdot10^{-6}$ \\
$\theta$ [°] & & & 27 & 36 & 34 \\
$k_0$ [$m^2$] & $1.72\cdot10^{-14}$ & $9.24\cdot10^{-14}$ & - & - & - \\
$\sigma'_{0}$ [$MPa$] & 350 & 495 & 350 & 300 & 325 \\
$n$ [$-$] & 11 & 2.5 & variable & variable & variable \\
$\rho_r$ [$kg\hspace{0.1cm}m^{-3}$] & - & 2400 & 2800 & 2200 & 2500 \\
$n_x, n_y, n_z$ [$-$] & - & [1,0,0] & [1,0,0]  & [1,0,0]  & [1,0,0]  \\
                                & & & [-0.555,0,0.832] & [-0.555,0,0.832] & [-0.555,0,0.832] \\
																& & & [0,0,1]  & [0,0,1]  & \\
$\lambda$ [$-$] & - & 0.41 & 1.5 & 1.5 & 1.5 \\
$Z$ [$m$] & - & 15 & - & - & - \\
\hline
\end{tabular}
}
\end{center}
\end{figure*}
\footnotesize{
\bibliographystyle{plain}
\bibliography{BiblioA1}}

\begin{thebibliography}{10}

\bibitem{Abaqus2008}
Abaqus.
\newblock {\em Abaqus theory manual, version 6.8}.
\newblock Systèmes Simulia Corp., Providence, R{I}, U{S}{A}, 2008.

\bibitem{Berkowitz2002}
B.~Berkowitz.
\newblock Characterizing flow and transport in fractured geological media: A
  review.
\newblock {\em Advances in Water Resources}, 25(8--12):861--884, 2002.

\bibitem{Brown1995}
S.R. Brown.
\newblock Simple mathematical model of a rough fracture.
\newblock {\em J. Geophys. Res}, 100(B4):5941--5952, 1995.

\bibitem{Cappa2006}
F.~Cappa.
\newblock Role of fluids in the hydromechanical behavior of heterogeneous
  fractured rocks: in situ characterization and numerical modelling.
\newblock {\em Bull. Eng. Geol. Env.}, 65:321--337, 2006.

\bibitem{COMSOLMultiphysics2010}
C{OMSOL}-Multiphysics.
\newblock {\em Multiphysics user's guide}.
\newblock COMSOL AB, Sweden, 2010.

\bibitem{Cornaton2007}
F.~J. Cornaton.
\newblock {\em Ground {W}ater: a 3-{D} {G}round {W}ater and {Surface} {W}ater
  {F}low, {M}ass {T}ransport and {H}eat {T}ransfer {F}inite {E}lement
  {S}imulator, {R}eference {M}anual}.
\newblock Centre for {H}ydrogeology and {G}eothermics, Neuchâtel,
  {S}witzerland, 2007.

\bibitem{Durham1997}
W.~B. Durham.
\newblock Laboratory observations of the hydraulic behavior of a permeable
  fracture from 3800 m depth in the {KTB} pilot hole.
\newblock {\em J. Geophys. Res.}, 102:18405--18416, 1997.

\bibitem{Ferronato2010}
M.~Ferronato, G.~Gambolati, C.~Janna, and P.~Teatini.
\newblock Geomechanical issues of anthropogenic {C}{O}{2} sequestration in
  exploited gas fields.
\newblock {\em Energy Conversion and Management}, 51(10):1918--1928, 2010.

\bibitem{Gangi1978}
A.~F. Gangi.
\newblock Variation of whole and fractured porous rock permeability with
  confining pressure.
\newblock {\em Int. J. Rock Mech. Min. Sci. Geomech. Abstr.}, 3:249--257, 1978.

\bibitem{Glover1998}
P.W.J. Glover, K.~Matsuki, Hikima R., and Hayashi K.
\newblock Synthetic rough fractures in rocks.
\newblock {\em J. Geophys. Res.}, 103(B5):9609--9620, 1998.

\bibitem{Hopkins2000}
D.L. Hopkins.
\newblock The implications of joint deformation in analyzing the properties and
  behavior of fractured rock masses, underground excavations and faults.
\newblock {\em Int. J. Rock Mech. Min. Sci. Geomech. Abstr.},
  37(1--2):175--202, 2000.

\bibitem{Itasca2006}
Itasca.
\newblock {\em {UDEC} Universal distinct element code}.
\newblock Itasca Consulting Group Inc., Minneapolis, U{S}{A}, 2006.

\bibitem{Kiraly1969a}
L.~Király.
\newblock Anisotropy and heterogeneity within jointed limestone.
\newblock {\em Eclogae Geologicae Helvetiae}, 62(2):613--619, 1969.

\bibitem{Kiraly1969b}
L.~Király.
\newblock Statistical analysis of fractures (orientation and density).
\newblock {\em International Journal of Earth Sciences}, 59(1):125--151, 1969.

\bibitem{Li2001}
S.~Q. Li, B.~Y. Xu, and Y.~G. Duan.
\newblock Coupling seepage of liquid and solid in fractured reservoir.
\newblock {\em Chin. J. Comp. Mech.}, 18(2):133--137, 2001.

\bibitem{Lombardi1988}
G.~Lombardi.
\newblock Les tassements exceptionnels au barrage de {Z}euzier.
\newblock {\em Publ. Swiss Soc. Soil Rock Mech.}, 118:39--47, 1988.

\bibitem{Lombardi1992}
G.~Lombardi.
\newblock The {FES} rock mass model - part one.
\newblock {\em Dam Engineering}, 3:201--221, 1992.

\bibitem{Londe1987}
P.~Londe.
\newblock The {M}alpasset dam failure.
\newblock {\em Eng. Geol.}, 24(1--4):295--329, 1987.

\bibitem{Louis1969}
C.~Louis.
\newblock A study of groundwater flow in jointed rock and its influence on the
  stability of rock masses.
\newblock Technical Report~9, Rock Mechanics, Imperial College, London, UK,
  1969.

\bibitem{Mayeur1999}
B.~Mayeur and D.~Fabre.
\newblock Measurement and modeling of natural stresses. application to the
  {M}aurienne-{A}mbin tunnel project.
\newblock {\em Bulletin of Engineering Geology and the Environment},
  58(1):45--59, 1999.

\bibitem{Murdoch2006}
L.C. Murdoch and L.N. Germanovich.
\newblock Analysis of a deformable fracture in permeable material.
\newblock {\em International Journal for Numerical and Analytical Methods in
  Geomechanics}, 30(6):529--561, 2006.

\bibitem{Neuman2005}
S.~P. Neuman.
\newblock Trends, prospects and challenges in quantifying flow and transport
  through fractured rocks.
\newblock {\em Hydrogeology Journal}, 13(1):124--147, 2005.

\bibitem{Rutqvist1996}
J.~Rutqvist and O.~Stephansson.
\newblock A cyclic hydraulic jacking test to determine the in situ stress
  normal to a fracture.
\newblock {\em Int. J. Rock Mech. Min. Sci. Geomech. Abstr}, 33(7):695--711,
  1996.

\bibitem{Rutqvist2002}
P.~Rutqvist, Y.-S. Wu, C.-F. Tsang, and G.~Bodvarsson.
\newblock A modeling approach for analysis of coupled multiphase fluid flow,
  heat transfer, and deformation in fractured porous rock.
\newblock {\em Int. J. Rock Mech. Min. Sci.}, 39(4):429--442, 2002.

\bibitem{Schweisinger2009}
T.~Schweisinger, E.J. Svenson, and L.C. Murdoch.
\newblock Introduction to hydromechanical well tests in fractured rock
  aquifers.
\newblock {\em Ground Water}, 47(1):69--79, 2009.

\bibitem{Terzaghi1923}
K.~Terzaghi.
\newblock Die berechnung der durchlässigkeitziffer des tones aus dem verlauf
  der hydrodynamischen spannungserscheinungen.
\newblock {\em Akad Wissensch Wien Sitzungsber Mathnaturwissensch Klasse IIa},
  142(3--4):125--138, 1923.

\bibitem{Tsang1981}
Y.W. Tsang and P.A. Witherspoon.
\newblock Hydromechanical behavior of a deformable rock fracture subject to
  normal stress.
\newblock {\em J. Geophys. Res.}, 86(B10):9287--9298, 1981.

\bibitem{Walsh1981}
J.~B. Walsh.
\newblock Effect of pore pressure and confining pressure on fracture
  permeability.
\newblock {\em Int. J. Rock Mech. Min. Sci. Geomech. Abstr.}, 18:429--435,
  1981.

\bibitem{ZSOIL}
Zace-Service-Ltd.
\newblock {\em Z-{S}oil.{PC}}.
\newblock Zace Service Consulting Group, Lausanne, Switzerland, 2010.

\bibitem{Zangerl2003}
C.~Zangerl, E.~Eberhardt, and S.~Loew.
\newblock Ground settlements above tunnels in fractured crystalline rock:
  numerical analysis of coupled hydromechanical mechanisms.
\newblock {\em Hydrogeology Journal}, 11:162--173, 2003.

\end{thebibliography}
\end{multicols}
\end{document}